# Comparison between Classical-Gas behaviours

# and Granular-Gas ones in micro-gravity :

# P. Evesque

Lab. MSSMat, UMR 8579 CNRS, Ecole Centrale Paris
92295 CHATENAY-MALABRY, France, e-mail: evesque@mssmat.ecp.fr

**Abstract:**

*Recent vibration experiments in microgravity have demonstrated that "granular-gas" state exists only in Knudsen regime, that its excitation is "supersonic" and that the probability density function of the pressure of the gas scales as* $(A\omega)^{3/2}$. *This paper pursues the description of the experimental results. Then, it compares these ones to what can be predicted from few simple modelling, which are (i) the classical-gas theory and (ii) the thermodynamics of a single particle in a* 1d *box. The anomalous scaling of pressure fluctuations* $(A\omega)^{3/2}$ *is explained by the crossover from the regime of a single collision during the sampling time, which imposes* p~ $(A\omega)$ *at small speed, to a multiple collision imposing* p~ $(A\omega)^2$ . *Air effect and effect of g-jitter are discussed and quantified. Influence of grain-grain collisions is described on the distribution of speed.*

**Pacs # :** 5.40 ; 45.70 ; 62.20 ; 83.70.Fn

_______________________________________________________________________________

Among the problems of granular matter, the granular-gas behaviour [1,2] is probably one of the most interesting. Its study consists in trying describing an ensemble of grains in chaotic fast motion via statistical mechanics and kinetic theory of gas, then to pass to the formalism of continuum fluid mechanics. One of the main goal is to identify which ones of the general results of the kinetic theory can be extended to granular gas and which ones cannot or have to be modified in order to take into account the energy losses generated during grain-grain collisions. Of course, one hopes that the general properties of the granular-gas medium can still be described using the same concepts as those used in the classical gas domain, *i.e.* temperature, pressure,….; but this remains an open question [3-5].

One of the most interesting properties of such "granular gases" is the tendency to form clusters [6,7]. Although this has probably been known since the early observation of planetary rings [8], only few recent laboratory experiments exist which display cluster formation [6,7,9]. Furthermore, ground based experiments [9] are perturbed by gravity and by the coherent friction force which acts on all the particles and which is far from being negligible. Indeed, low gravity environment is quite needed in this domain of research because it is the only way to achieve the experimental situation in which inelastic collisions are the only interaction mechanism.

In a previous experiment on board of the CNES micro-g Caravelle[1], the dynamics of few grains in a large vibrated container (amplitude A, pulsation $\omega=2\pi f$) was investigated and it was found that these dynamics obey the Boltzmann statistics at least approximately; this experiment tends then to prove that the concept of granular-gas is an efficient concept. In a more recent experiment [6,7] however, the density of grains has been increased. These experiments have demonstrated that the inelasticity of the collisions provokes the formation of a large cluster as soon as the mean free path $l_c$ between two grain-grain collisions is larger than the cell size; hence this limits strongly the range of observation of the granular-gas phase and demonstrates the existence of a transition from a "gas"-phase to a more condensed phase. Furthermore the data reported in [6] on the density function of pressure fluctuations, have found that the granular-pressure p of the "gas"-phase in the vicinity of the "gas"-cluster transition, *i.e.* $l_c \approx L$, scales with the typical box speed $V=A\omega$ as $p \sim (A\omega)^{3/2}$ instead of what one should expect from a dimensional analysis, *i.e.* $p \sim (A\omega)^2$. Then, in [7], it was demonstrated that the excitation due to the wall looks supersonic since a depletion zone exists near the wall where no grain can be found when the wall moves backward. So, these experiments [6,7] tend to demonstrate that the statistics of grain velocity does not obey the classical Boltzmann distribution when grain density becomes large.

In this paper, the analysis of experimental data of ref. [6,7] is summed up and pursued in the first section. The second part recall what shall be expected from a perfect-gas analogy and the different kinetic regimes that are expected in that case. The third section discusses the experimental data and compares them to the models of perfect gas and to the single-particle model, which considers that grains do not interact together, but interact only with the boundary [10].

| Experiment number | Time segment (s) | Amplitude $A$ (mm) | Frequency $f$ (Hz) | Velocity $V$ (cm/s) | Acceleration $G$(g unit) |
|---|---|---|---|---|---|
| 1 | 23 - 36 | 0.1 | 3 | 0.2 | 0.004 |
| 2 | 46.8 - 52 | 2.5 | 1 | 1.6 | 0.01 |
| 3 | 52.3 - 67.3 | 2.5 | 3 | 4.7 | 0.09 |
| 4 | 76.5 - 84.5 | 0.3 | 30 | 5.6 | 1.1 |
| 5 | 87 - 100 | 0.1 | 60 | 3.8 | 1.4 |
| 6 | 103 - 116.5 | 0.3 | 60 | 11.3 | 4.3 |
| 7 | 120 - 130 | 1 | 60 | 37.7 | 14.5 |
| 8 | 138.5 - 148 | 2.5 | 30 | 47.1 | 9 |
| 9 | 151 - 180 | 2.5 | 60 | 94.2 | 36.2 |

*Table 1: Vibrational parameters during the 200 s of low gravity. Time segment is the duration of each experiment at fixed amplitude and frequency without taking into account the transient states.* $V = 2\pi Af$ and $\Gamma = 4\pi^2 Af^2/g$ *are respectively the velocity amplitude and the dimensionless acceleration amplitude of the vessels,* g=9.81m/s². *Data from experiment #1-2 had a very poor signal to noise ratio.*

## 1. Results of MiniTexus 5 experiments on vibrated granular gas in micro-gravity

### *1.1. Experiment*

The Mini-Texus 5 space-probe was launched on 11/2/1998 from Esrange (Nothern Sweden) on a Nike-Improved Orion rocket with 3 cubic containers on board, 1 cm³ in inner volume, with clear sapphire walls. Each cell is filled, respectively, with 0.281, 0.562 and 0.8915 g of 0.3-0.4 mm in diameter bronze spheres (solid fractions: 3.2%, 6.4% and 10.1%). Thus, the total number of particles in each cell is about 1420, 2840 and 4510, corresponding to roughly 1, 2 and 3 particle layers at rest. An electric motor, with eccentric transformer from rotational to translation motion, drives the vessels sinusoidaly at frequency *f* and maximum-displacement amplitude *A* in the ranges 1 to 60 Hz and 0.1 to 2.5 mm respectively. The vibration parameters during the time line are listed in Table 1 [6]. The vessel containing the three cells filled with bronze beads is displayed in Fig. 1 of [7] together with the three pressure gauges.

Motion of particles was visualised and recorded by a SWMO39 CCD camera, fixed in the frame of the space-probe, that captures 742 x 582 pixel images with a 40 ms exposure time. Each cell was illuminated by one led lighting in the Ox direction at right angle from the Oy observation. Direction of mechanical excitation is Oz. Continuous illumination was used when mechanical-excitation frequency was smaller than 30Hz, and stroboscopic illumination of 1ms duration and 14 Hz (*resp.* 28 Hz) repetition rate was used when the driving mechanical frequencies was 30 Hz (*resp.* 60Hz). $(\Gamma_x, \Gamma_y, \Gamma_z)$ accelerations were measured by piezoelectric accelerometers (PCB 356AO8) screwed in the shaft in a triaxial way. Typical output sensitivities in the vibration direction and in the perpendicular directions were, respectively, 0.1 and 1 V/g, g =9.81 m/s² being the acceleration of gravity. A piezoelectric pressure sensor (PCB 106B50), 1.53 cm in diameter, was fixed on the "top" of each cell to measure "grain pressure" $p_z$ due to momentum transfer. The accelerometer orientation, in the direction of vibration, was such that its head was pointed perpendicular towards pressure sensor surface. Typical pressure sensor characteristics were a 0.72 mV/Pa output sensitivity, a 20 Pa/g acceleration sensitivity, a 40 kHz resonant frequency, and a 8 μs rise time. $(\Gamma_x, \Gamma_y, \Gamma_z, p_z)$-data acquisition was performed with a 2 kHz sampling rate during the 200 seconds of low gravity environment (about $10^{-5}$ g) and were transferred to Earth in real time. The firing of the engine, the stabilisation of the rocket on its parabolic trajectory and the despinning of the rocket lasted roughly 90 s; then, the first 20 s of the experiment was without vibration to let the granular medium relaxing.

## *1.2. Results and discussion:*

### 1.2.a. video analysis:

*probing cluster formation:* Fig.1 displays what can be observed in the three different cells when excited at *f* =60Hz , with *A*=1 mm, and at two different phases of the vibration cycle: Fig. 1.a (*resp.* Fig. 1.b) corresponds to "upward" (*resp.* "downward") velocity; an other example, corresponding to experiment # 8, *i.e. f* =30Hz, *A*=2.5 mm, is given in Fig. 1 of [6] . The density is decreasing from the left to the right. In the most dilute case, the particles move erratically and their distribution is roughly homogeneous in space (there is a depletion close to the boundary moving away from the particles). In the two denser cases, a motionless dense cluster in the reference frame of the camera, *i.e.* of the space-probe (black central region of the photographs) is surrounded by lower particle density regions. The spheres surrounding the cluster are in motion. We thus observe that at high enough density, the spatially homogeneous gas of particles undergoes an instability which leads to the formation of a dense cluster. Note that the left and the middle (*resp.* right) cells are illuminated from the left (*resp.* right) side.

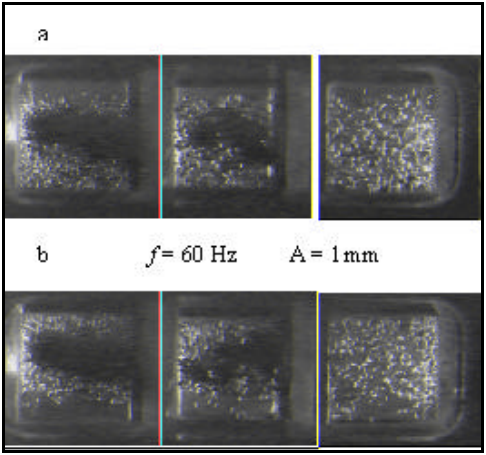

The analysis of the light intensity can be performed from a statistical analysis: let $n(x,y,z)$ be the particle density in the cell and let divide the cell volume in elementary volume having the size of a grain $a^3$; the light intensity $I(x_o,z_o)$ observed at location $(x_o,z_o)$ is due to light diffusion by a particle at $(x_o,y,z_o)$ which is not intercepted by other beads. So it can be written as:

$$I(x_o,z_o)a^2 = I_o a^2 \sum_{y_1=0\,to\,L/a} [\Pi_{o\text{-to-}x_o/a}\{1-n(x,y_1,z_o)a^3\} \; \Pi_{o\text{-to-}y_1/a}\{1-n(x_o,y,z_o)a^3\} \; n(x_o,y_1,z_o)a^3]$$

Products in each $\Pi$ corresponds to the probability that the light beam does not interact with a grain contained in an elementary volume located on the beam trajectory; the last $n(x_o,y_1,z_o)$ term corresponds to the presence of a scattering grain at location $(x_o,y_1,z_o)$. The summation $\Sigma$ takes into account all the possible position $(x_o,y_1,z_o)$ of the diffusing grain and all the possible beam trajectories corresponding to $I(x_o,z_o)$.

$$I(x_o,z_o) = I_o \int_0^L a^2 n(x_o,y,z_o)\, dy \, \exp[-a^2 \int_0^{x_o} dx\, n(x,y,z_o) - a^2 \int_0^y du\, n(x_o,u,z_o)] \tag{1}$$

When $n(x,y,z)$ depends on z only, Eq. (1) can be rewritten:

$$I(x_o,z_o) = I_o \int_0^L a^2 n(z_o)\, dy \, \exp[-a^2(x_o+y)\, n(z_o)]$$

$$I(x,z)/I_o = \exp[-a^2 x\, n(z)] - \exp[-a^2(x+L)\, n(z)] \tag{2.a}$$

$$I(x,z)/I_o = \exp[-a^2 x\, n(z)] \, \{1 - \exp[-a^2 L\, n(z)]\} \tag{2.b}$$

So, Eq. (2.b) explains , at least qualitatively, the apparent increase in cluster size for the two left cells: it is an artefact due to light diffusion which reduces, as $\exp[-a^2 xn]$, the enlightenment as the light penetrates deeper in the cell, *cf.* Fig. 1. In the same way,

Eq. (2.b) explains why the edge of the cluster is brighter than the external part; it is due to the {1-exp[-na²L]} term which is small , *i.e.* ≈na²L, outside the cluster and equal to 1 about inside the cluster.

♣ *"Supersonic" excitation of the "gaseous" phase:* One observes in the three cells of Fig. 1 the existence of a sharp discontinuity of the bead density just nearby the upper (*resp.* lower) wall, *cf.* Fig. 1.a (*resp.* 2.b). It means that the "granular-gas" is also surrounded by a zone where no grain is present; and the limit of these black zones corresponds, at least approximately, to the minimum (*resp.* maximum) height of excursion of the top (*resp.* bottom) wall of the cells. It means that the typical speed of a grain pertaining to the "granular-gas" phase is much smaller than the typical cell speed, *i.e.* Aω. This demonstrates the supersonic nature of the cell motion compared to the "granular-gas" temperature. One can estimate the relative speed of the cell in gas-speed unit by noting that the time needed for the expansion burst to reach the wall is approximately 3/(4f), where f is the frequency of the mechanical excitation. This leads to evaluate the cell speed to be Mach 3-4 about, in the unit of "granular-gas" speed. All these facts are confirmed by the other observations at the different frequencies and amplitudes, *cf.* Fig. 1 in [6] and Figs. 2 & 4 in [7].

One notes also that the cluster of the two denser cells is not excited by the moving wall directly, but via the "granular-gas" phase, which plays the role of an interface.

♣ *Knudsen regime of the "gaseous" phase:* The "granular-gas" phase of each cell is very loose since one sees each grain separately. It implies that the mean free path $l_c$ between two grain-grain collisions is larger than the cell size L. It means then that the "granular-gas" phase is in the so-called Knudsen regime [11]. This regime is quite specific since one can define local averages such as density ρ, pressure p, temperature T which have some meaning according to the ergodic principle; but these variables cannot be used to predict the sample evolution using classical hydrodynamics which assumes that the evolution of each elementary volume depends only on adjacent elementary volumes. On the contrary one shall introduce non local interaction here. This is why it is difficult to speak in term of wave propagation and of hydrodynamics. This is why we label this phase the "granular-gas" with inverted commas.

## 1.2.b. Signals from gauges

Fig. 2 displays $\Gamma_z$ together with the total granular pressure signal $p_z$ of the loosest sample during the same experiments (#5 - #9). The third curve of each Figure corresponds to the $\Gamma_z \bullet p_z$ correlation function, *i.e.* $\Gamma_z \bullet p_z = \int d\tau \ \Gamma_z(t) \ p_z(t+\tau)$. Fourier Transform of $\Gamma_z$ acceleration signals are given in Fig. 3 in the case of experiments #7,8,9; it proves that the excitation is rather monochromatic except for exp. #9 (*i.e.* f=60Hz, a=2.5mm) for which some harmonics are observed.

The difference between the two kinetic regimes, homogeneous and clustered, is also apparent on the pressure signals. Typical examples are displayed in Fig. 4, *i.e.* Fig. 2 of [6]. The part of the pressure signal which is due to the gauge sensitivity to acceleration has been removed by signal processing using Fourier transform.

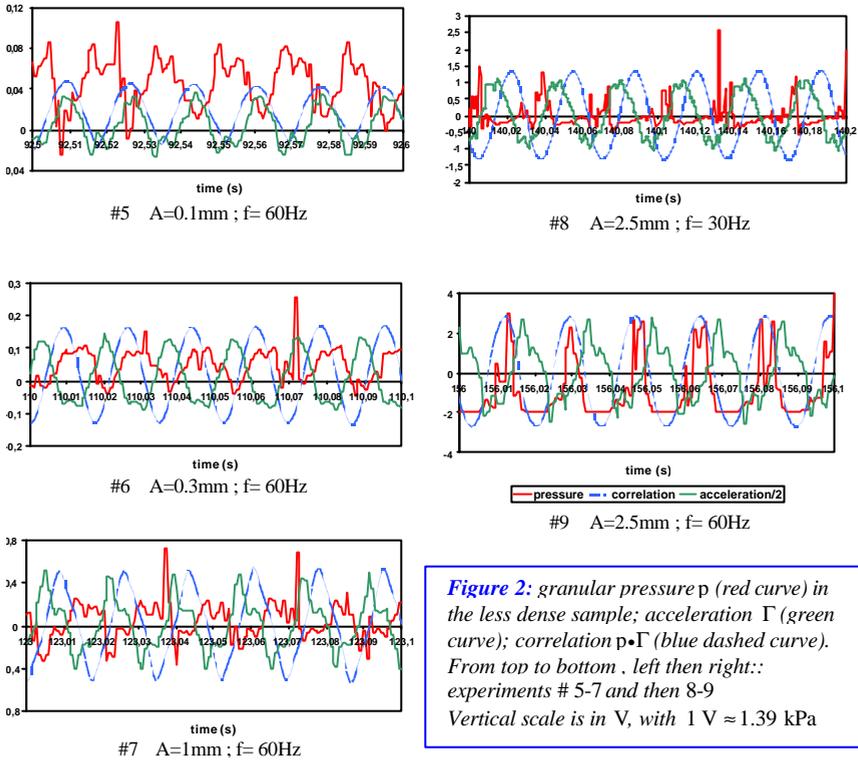

In the dilute case (upper curve in Fig. 4), the time recording of the pressure, measured at the "top" wall, shows a succession of peaks corresponding to particle collisions with the wall. In the two denser cases, the pressure involves a component in phase with the acceleration imposed to the vessel (lower curve in Fig. 4). Note that in the case of intermediate density (second signal from the top), both the pressure peaks and the component in phase with the acceleration are visible. However, the amplitude of pressure fluctuations is smaller than for the dilute case, although the particle number is larger; the reason is that most particles are in the cluster, which, as can be seen in the video recordings, stays away from the walls. The pressure component, in phase with the acceleration, traces back to the grains in the low-density region between the cluster and the walls. It shows that in the densest case, the motion of these particles is coherent with the vibration. In the case of intermediate density, the vibration generates both a coherent pressure oscillation and incoherent motions displayed by the random pressure peaks in the signal.

However as the video demonstrates there is no grain in contact with the pressure sensor during some part of the period, one shall then admit that this sinusoidal signal is generated via the air pressure most likely.

♣ Dilute "granular-gas" case: We now consider the dilute case for which the spatially homogeneous fluidised regime is stable. Particles move erratically and the pressure signal displays a succession of peaks. Bursts of peaks occur roughly at $\pi/2$ with the acceleration but the number of peaks in each burst, their amplitude, and the duration of each burst are random (see Fig. 4). Note the small distortions in the acceleration signal, occurring at the same times as the pressure peaks; these distortions near the extrema of the acceleration signal are generated by the mechanical driver at full stroke. Let us estimate that peaks cover periodically 1/10 of a period; be $\tau_s$=0.5ms the sampling time; so one can estimate that each peak of pressure recorded in the upper curve of Fig. 4 involves about 4 $N_o$ Af $\tau_s$/(2L) < 20 collisions.

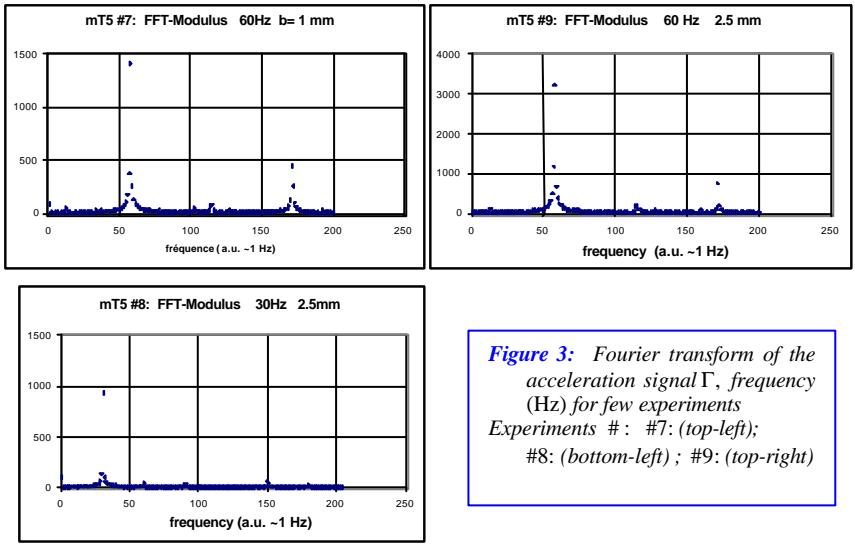

**Figure 3:** *Fourier transform of the acceleration signal* $\Gamma$, *frequency* (Hz) *for few experiments Experiments # : #7: (top-left); #8: (bottom-left) ; #9: (top-right)*

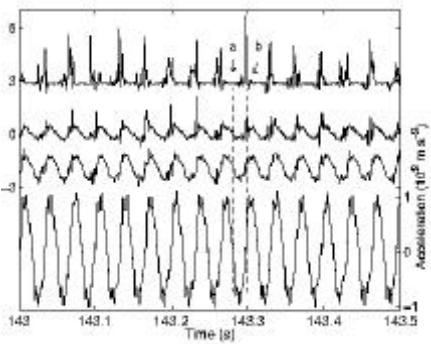

**Figure 4:** *Granular pressure variations as a function of time (s) for the three cells, for experiment # 8 in Table 1 .*
*From the upper to the lower curve: pressure signal in the dilute sample; pressure signal in the medium dense sample; pressure signal in the densest sample; acceleration in the direction of vibration. Pressure curves are shifted vertically for clarity.*
*Spikes of top curve demonstrate the incoherent motion of the granular gas in the dilute regime. However each spike correspond to a series of grain-wall collisions.*

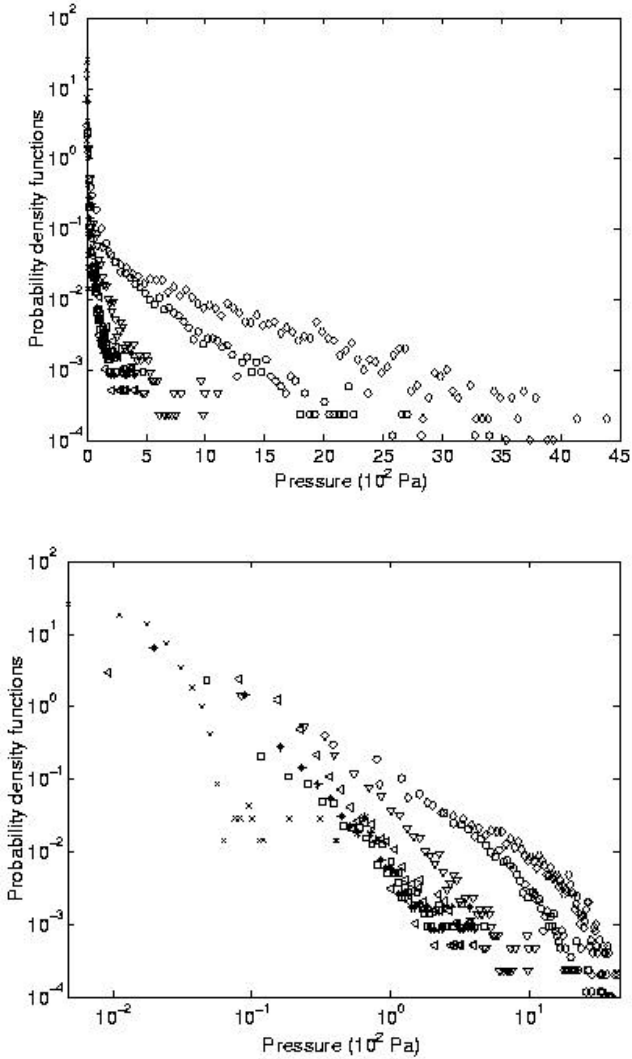

**Figure 5:**

*Probability density functions of pressure fluctuations in the dilute cell, for the different vibration parameters of Table 1.*
*5.a Semi-log plots.*
*5.b Log-log plots*

*Experiment # :*
  #2 (x);  #3 (**O**);
  #4 (*);  #5 (◁);
  #6 (¬); #7 (○);
  #8 (◊).

♣ Pressure in the dilute cell: As above mentioned already, one can obtain from Fig. 4 the proportion of time during which particles are in contact with the sensor in the dilute cell. It is roughly T/4. This signal displays large fluctuations whose statistics is reported in Fig. 5 in the form of the probability density functions of pressure fluctuations, measured directly by the pressure sensor. Rescaling procedure of these data was performed in [6] to obtain a single curve, indicating that the probability density functions roughly scales as $V^{3/2} = (2\pi A f)^{3/2}$, where V is the maximal vibration velocity of the vessel; V ranges from 1.6 to 47 cm/s.

## 2. Classical gas of atoms in a vibrated box:

Let us first consider the case of a vibrating box of size $L^3$ filled with a gas of atoms at pressure p and temperature T. Be $z=A \cos(\omega t)=A \cos(2\pi f t)$ the motion of the box. One can classify the different regimes according to natural length and speed. So be $l_c \approx \mu/(\pi d^2 N_{oA} \rho)$ the mean free path of a molecule and c the speed of sound in the gas, *i.e.* $c=[(\partial p/\partial \rho)_s]^{1/2}=(\gamma \Re T/\mu)^{1/2}$, where $\Re=N_A k_B$ is the gas constant, $N_A=6.02 \ 10^{23}$ the Avogadro number, $\gamma=C_p/C_v$, $\rho$ the gas density, $k_B$ the Boltzmann constant, $\mu$ the molecular mass of the gas; so $\mu_a=\mu/N_A$ is the mass of an atom. Be also $\lambda=c/f$ the wavelength of the sound wave propagating in the gas, L the box size and $V=2pAf$ the speed of the box. One has to classify the different regimes as functions of $l_c, L, \lambda$ and of V and c:

**Influence of the mean free path $l_c$:** Applying the sound wave formalism (or any hydrodynamics formalism) requires that the system be much larger than the mean free path $l_c \approx \mu/(\pi d^2 N_A \rho)$: Indeed, this formalism describes the evolution of an elementary volume via local coupling with nearest neighbours only. Hence this requires to consider elementary volumes having a size larger than $l_c$. Applying such a condition to sound wave and considering a gas of mono-atomic molecules of diameter d with a specific volume per molecule $n =\rho/\mu$ lead to the condition on the wavelength: $\lambda > l_c$ . Indeed, this is a known result: sound transmission does not exist when $\lambda < l_c$ [11,12,15,16].

**Knudsen regime:** Then different cases have to be considered depending on the relative value of L , $\lambda$ and $l_c$. They are summed up in Table 2. If $L < l_c$, the problem falls in the microscopic scale and the gas is known to be in the Knudsen regime [11]; this seems to be the case for the "granular-gas" phase reported in Section 1. However, it is worth noting that sound propagation has been studied in very-low-pressure gases [15,16] even when $L < l_c$ [15,16] using two parallel plates at temperature T; indeed, the statistics of the response to a vibrational excitation of a plate can be calculated in this microscopic regime and compared to experiments. It results from the experimental condition that most of the atoms transfer directly their momentum to the walls, and that some of them only collide together.

On the contrary, if $L >> l_c$, the gas obeys classical laws; and one has to consider whether $\lambda > L$, or not, and whether $V=A\omega < c$ or $A\omega > c$.

**Subsonic excitation:** sound propagates in condition of subsonic excitation, *i.e.* $A\omega < c$, otherwise one gets a shock wave, *cf.* [14]. So, when $a\omega < c$, one can develop a wave-propagation formalism. This requires to neglect the term $\rho v.\nabla v$ in front of $\partial \rho/\partial t$. This imposes in turn $\delta \rho/\rho << 1$ or $v << c$ , where $\delta \rho$ is the variation of $\rho$ due to the wall motion; as one expects v=Aω near the wall, this imposes A<<λ=c/f. The wave-propagation formalism can neglect temperature-diffusion effect and viscous force when the amplitude of pressure wave is small compared to the mean pressure; otherwise it cannot, which leads to absorption and dephasing [14]. We now neglect these effects and consider only a case when δp/ρ<<1.

*Subsonic excitation, L<λ*: When L<λ, the gas can be considered as incompressible. The cell vibration imposes a pressure variation δp in the gas which keeps a slice of gas in equilibrium with its neighbours; so δp depends on z and t according to δp = -ρ z A ω² cos(ωt) at position (x,y,z) of the cell, where ρ=μp/($\mathfrak{R}$T).

Integration of p along the sensor size leads to the measured pressure variations: For instance, let us consider a pressure gauge which occupies the whole surface of one side of the cubic box, the pressure variation will be zero when the pressure-sensor is parallel to the vibration; on the contrary, when its normal is parallel to the vibration direction, it is δF =δp L² = -ρ L³/2 A ω² cos(ωt).

*Subsonic excitation, L>λ*: when L>λ/2, the sound propagates in the cell; its amplitude depends on resonance conditions, *i.e.* L=k c/(2f), with k being any positive Integer. Amplitude of sound waves at resonance are limited by dissipation terms such as viscous and thermal-diffusion effects [14], which will not be described here. Anyhow, it is worth noting that the kinematic viscosity of a gas is of order $\nu=cl_c$.

**Supersonic excitation and shock waves:** when Aω becomes larger than the speed of sound c, *i.e.* Aω>c , the excitation shall produce a shock wave which is a surface of discontinuity between the two parts of the gas; this surface propagates with the speed of sound. The width δ of the discontinuity surface is controlled by temperature-diffusion processes and viscous effects [14]. It is of order δ≈$l_c$ when relative pressure difference is about Δp/p=1; it increases when Δp/p decreases. One shall know also that discontinuity surfaces are only stable when the discontinuity moves forward into the low pressure phase [14].

| | **L >> $l_c$** | | **L<$l_c$?** |
|---|---|---|---|
| | L>λ>$l_c$ | λ> L>$l_c$ | $l_c$ >L |
| aω<c $c\approx(k_BT/\mu_a)^{1/2}$ | Subsonic compressible gas | Subsonic "incompressible" | "Subsonic" Knudsen regime |
| aω>c $c\approx(k_BT/\mu_a)^{1/2}$ | Supersonic non linear behaviour | Supersonic non linear behaviour | "Supersonic" Knudsen regime |

**Table 2:** *The different cases of vibrated excitation of a gas of atoms at temperature T*

*Depletion zone:* So when the cell moves forward in the gas it generates a compression shock waves which propagates towards the other wall. But when it moves backward the discontinuity surface is not stable and the gas expands adiabatically in the direction of the moving wall, with a typical speed that is equal to the sound speed c. So, during the back motion the gas expands; it is reconfined at the next period. As the gas expands, and if A>$l_c$, the gas cools down periodically, leading to temperature- and density- inhomogeneity. As the medium cannot expand as fast as the wall motion, the gas remains confined in a volume $\xi$ smaller than the cell size L, *i.e.* $\xi \approx$ L-2A, and the wall "hits" periodically the gas when the cell position is at full stroke.

When the cell size L is such as L>c/f, more than one shock wave can be observed at the same time propagating in the cell in both $\varepsilon z$ directions, $\varepsilon = \pm 1$; they are generated periodically at each wall at times $t_n=1/(4f)[1-\varepsilon]+n/f$. Conditions of resonance can be also obtained for special frequencies f, such as c/f=L .

When the cell size is L<c/(2f ), only one shock wave can be observed at the same time propagating in +z or -z directions.

### *Few remarks:*

Remark concerning the case of Subsonic excitation with L<$\lambda$: One can try and find the incompressible subsonic result from statistical mechanics consideration; but this requires a minimum of care and to introduce $l_c$. Let us show a wrong demonstration. We start from the Boltzmann distribution of speed u, $\rho(u)\sim$exp-$(u^2/(2k_BT)$. The pressure p=$\delta F/(\delta S \delta t)$ is the force per surface area and per time unit which is generated due to the hitting of atoms on the sensor membrane; it is then proportional to the total transfer of momentum during $\delta t$. As this one is equal to the product of the momentum transfer $\mu_a(v+u)$ of a particle at speed u hitting the wall at speed v, by the number of collisions during $\delta t$, *i.e.* (v+u)$\rho$(u) du, so p $=\sum_{coll} 2m(v+u)(v+u)\rho(u)$ du. As $\rho(u)$ is symmetric, integration leads to variations p $\approx$v². *So, this is not the correct answer.*

Indeed the correct value, *i.e.* $\delta p = -\rho z A \omega^2 \cos(\omega t)$, is obtained from the continuum-mechanics approach described above. The statistical approach proposed here is not valid for the following reason: it does not take into account the fact that the medium is in mechanical equilibrium. Indeed, in real case, the gas reacts at once when the wall moves; this leads to a compression of the gas nearby the wall, and this compression does *propagate* in the gas at the speed of sound. This modifies the pressure distribution in the cell at once; the pressure is no more the initial pressure, as it was supposed in the previous paragraph.

Remark about the Representative Elementary Volume (REV) of a gas: It is worth noting that the REV $\varpi_o$ above which classical hydrodynamics equations are valid is defined by $\varpi_o \approx l_c^3$. Since $l_c \approx \mu/(\pi d^2 N_A \rho)=1/(n \pi d^2)$, where n is the particle density, the REV $\varpi_o$ scales as $\varpi_o \approx 1/(\pi^3 d^6 n^3)$ , which contains in mean $N_\varpi \approx 1/(\pi^3 d^6 n^2)$ molecules. As $N_\varpi$ increases as the square of the mean volume per molecule, so the number of particles in the REV increases strongly when the system is rarefied as in a gas. On the contrary, one gets $N_\varpi \approx 1$ when the system is dense as in the case of a liquid or of a solid. This effect seems to be completely ignored from the community of physicists working in the granular-matter area, who assumes quite often that small REV is the requirement to get classical physics.

This REV size does not imply that $pv = n\Re T$ is a law which cannot be observed in cells smaller than $l_c^3$. Indeed, the $pv=n\Re T$ law is observed in mean when $L<l_c$, (and even for a single molecule), due to the ergodic principle; but one has to be aware of long-range correlation which makes the equation of continuum media more complex.

Remark concerning subsonic excitation in the limit of large $a\omega$: $a\omega\approx c$, *but* $a\omega<c$: *sound wave propagation and acoustic streaming...* When sound-wave-resonance occurs, the problem becomes more complex because acoustic streaming can be generated, which produces in turn *new dissipation process* and vortex-flow [17-20]. Indeed the mechanism that generates these flows is linked to the generation of averaged vorticity in inhomogeneous viscous skin layers [17-20]. It has been shown that these flows are controlled by at least two parameters: first, they depend on the ratio of the vibration amplitude A to the thickness of the viscous skin-layer or, what is about the same, on the Reynolds number $R_e=A^2\omega/v$; second, they depend also on the size of the cell through the dimensionless parameter $\Omega$, *i.e.* $\Omega=\omega L^2/v$, [17-20].

Remark concerning Pressure fluctuations: If N is the number of collisions generating the pressure variations, the pressure p is given by $pS\delta t=Nm\underline{v}=mv\,n\,\underline{v}\,S\delta t$, where S is the gauge surface, $\delta t$ the integration time and n the particle density, $\underline{v}$ being the mean velocity. When all particles move independently from one another, the statistical fluctuations S $\Delta p$ of F=Sp scales as $N^{1/2}\delta(mv)/\delta t$, where $\delta(mv)$ is the typical fluctuation of particle momentum. As, for a gas $\delta(mv)\approx m\underline{v}$, one gets

$$F\delta t=pS\delta t\approx m\underline{v}^2n\delta t \qquad \text{and} \qquad \delta(pS\delta t)\approx m\underline{v}(\underline{v}Sn\delta t)^{1/2}.$$

When one works with a fixed gauge of surface S, with a given density of particles and a given observation time $\delta t$, the previous equations predicts that the typical fluctuations of pressure $\Delta pS\delta t$ shall scale as $\underline{v}^{3/2}$ according to:

$$\Delta(pS\delta t) \approx m\,\underline{v}^{3/2}\,n^{1/2}\,(S\delta t)^{1/2} \tag{3.a}$$

$$p_oS\delta t =F\delta t \approx m\,\underline{v}^2\,n\,\delta t \tag{3.b}$$

Another point which is worth mentioning is the following: as far as the duration $\delta t$ and the surface S are large enough, the number of collisions per $\delta t$ is much larger than 1. In this case, one expects that the density distribution of the signal is Gaussian, *i.e.* $\rho(p)\sim\exp\{-(p-p_o)^2/(\Delta p)^2\}$; it is distributed around its mean value $p_o$ which is given by Eq. (3.b), and its typical width $\Delta p$ is given by Eq. (3.a). This is a consequence of the central limit theorem and of the fact that the motions of the molecules are not correlated.

Indeed, the Maxwellian distribution, predicted by the kinetics theory of gas, *i.e.* $\rho(v_z) \sim \exp[-mv_z^2/(2kT)]$, is only observable when the number of collisions during $\delta t$ is smaller than 1. In this case the pressure is proportional to $mv_z$ and the pressure distribution shall then scale as $\rho(p) \sim\exp[-p^2/(2mkT)]$. Hence it is still a Gaussian, but it is centred on 0. So, in the case of perfect gases of atoms or molecules, $\rho(p)$ is always Gaussian-shaped; it is then a general feature, which is independent of $l_c$ and of the density of the gas; it applies in the Knudsen regime as well as in a denser case. Only the mean $p_o$ and the width $\Delta p$ of the Gaussian do depend on the size of the measuring surface and on the integration time. Moreover $p_o$ and $\Delta p$ do not depend on $l_c$.

These results are important since they shall serve as a reference to understand the scaling law of the density fluctuations of the pressure signal in the case of a "granular-gas" experiment. Indeed, it is quite surprising that the $\rho(p)$ of the "granular-gas" experiments of Fig. 5 follows different exponential shape, *i.e.* $\rho(p)\sim\exp(-p/p_1)$ and are never Gaussian-shaped.

### 3. Discussion & Conclusion:

#### *3.1. Few questions:*

♣ Mean pressure, distribution of pressure fluctuations:
Last paragraph of the previous section has pointed out one of the main interrogation these experimental results lead to: why do the statistics of the pressure distribution is not Gaussian-shaped? This makes the behaviour of a granular-gas quite different from the classical-gas behaviour. An other problem with the experimental data on pressure fluctuations is that they lead to a scaling $p\sim(A\omega)^{3/2}$ instead of the theoretical prediction $p\sim(A\omega)^2$. Indeed this theoretical prediction $p\sim(A\omega)^2$ can be obtained either from a direct argument of dimension analysis, or from an argument on the total transfer of momentum of a collection of beads having typical speed.

Are we able to understand these two incompatibilities? In a previous work [6], the cause of the $p\sim(A\omega)^{3/2}$ has been attributed to the existence of a restitution coefficient which varies with v. Is it the only possible explanation?

#### *3.2. A new possible explanation for the* $p\sim(A\omega)^{3/2}$ *scaling:*

♣ Analogy with a single particle model:
Indeed, other analogies can be proposed. In particular, one can think that particle-particle interaction is negligible due to the very dilute condition when homogeneous granular gas is observed. So the problem can be mapped on the model of a single particle in a vibrated container. Starting from a 1d analogy [10], one finds that the speed distribution is peaked around a mean velocity $v_o$, with a width $\Delta v_o \approx v_o/3$ in general, *i.e.* when A/L<0.005. This is the case when A/L<0.005 and/or when the restitution coefficient r (defined as $[v'_1-v'_2]=-r[v_1-v_2]$) is smaller than 0.95.

In this case, it was found [10] that $A\omega<v_o$ and that $v_o$ scales as :

$$v_o\approx A\omega\,(1+\varepsilon)^{1/2}\,/[2(1-\varepsilon)]^{1/2} \qquad (4)$$

When resonance occurs, which implies large A/L>0.005 and large r>0.95 [10], simulations find a sharpening of the speed distribution around a set of precise values; it is associated to a strong increase of the correlation time [10] which is characteristic of "non-ergodic" behaviour.

As the problem has a natural scale, L, $v_o$ can depend also on A/L. This is true only at large values of A/L, *i.e.* A/L>0.01 [10]; this last effect is equivalent to a finite size effect. In particular, computer simulations at constant $\omega$ find a decrease of the ratio $v_o/(A\omega)$ when A increases in the range 0.01<A/L< ½, which is the experimental range used in the MiniTexus-5 experiments. This may then explain the experimental p~$A^{3/2}$ scaling, but not the $\omega^{3/2}$ experimental term.

Another question arises: in the 1-particle model, one gets a mean speed $v_o$ which is always larger than A$\omega$. So why do we observe always smaller values of $v_o$ in experiments: $v_o \approx 1.5Af \approx A\omega/4$ ? Is it due to a 3d effect?, or to the coupling between translation and rotation degrees of freedom? We will come back later on this problem; it is first needed to precise the real experimental conditions.

### *3.3. Few orders of magnitudes:*

#### ♣ Number of counts per sampling time, typical pressure signal

Consider the most dilute cell; be $N_r$ the sampling rate ($N_r$=2kHz); be $\tau_s$=$N_r^{-1}$ the sampling time; be $N_o$=1420 the number of particles of mass m=0.28/1400=0.2mg and of radius R=0.15-0.2mm in the cell of size $L^3$ , (L=1cm). According to Fig. 1, at each period the granular cloud expands over the amplitude A about on both sides, leading to a number $N_c$ of collisions per period given by $N_c \approx AN_o/(2L)$. According to Figs. 1 & 4, these collisions occur during ¼ of the period about. It means that the number $n_c$ of collisions for each sampling channel is $n_c \approx 2AN_of/(N_rL)$=1.4 Af/L , (where 1.4 is in s). Taking into account the fact that the particles move at a typical speed Af slowlier than the gauge one (2$\pi$Af), one can estimate the typical maximum momentum-transfer m$\Delta$v during a single collision; this leads to a maximum force $\delta F_{cm}$ =m$\Delta$v/$\tau_s$ per collision given by $\delta F_{cm}/N_r$= m[2(2$\pi$+1)Af]; this leads in turn to a maximum pressure-increment at each collision $\delta p_{cm}$=$\delta F_c/S_o$= 2m(2$\pi$+1)AfN_r/S_o, where $S_o$=1.84cm² is the gauge surface. So, $\delta p_c$=32Af, where 32 is 32 Pa s/m.

Table 3 reports the expected values of $n_c$ , $\delta p_c$ and of the typical mean pressure p=$n_c\delta p_c$ for the different experimental cases.

| Exp # | 1 | 2 | 3 | 4 | 5 | 6 | 7 | 8 | 9 |
|---|---|---|---|---|---|---|---|---|---|
| $n_c$ | 0.04 | 0.35 | 1 | 1.26 | 0.84 | 2.5 | 8.4 | 10.5 | 21 |
| $\delta p_c$ | 0.01 | 0.08 | 0.24 | 0.29 | 0.19 | 0.58 | 1.9 | 2.4 | 4.8 |
| p=$n_c \delta p_c$ | 0.0004 | 0.03 | 0.24 | 0.37 | 0.16 | 1.5 | 16 | 25 | 101 |
| losses | Large viscous | Large viscous | viscous | viscous | viscous | viscous | viscous | viscous | collisional |

***Table 3:*** *Typical number* $n_c$ *of collisions per acquisition time* (0.5ms), *typical mean expected pressure* $p_o$ *and typical losses ( collisional vs.viscous) as a function of the experiment number.*

### 3.4. Effect of the ambient air:

♣ Effect of the air viscosity on the dynamics of a single grain:

Air applies a viscous drag force $F_v$ on each grain of radius $R \approx 0.15mm$ and mass $m \approx 0.2mg$ according to $F_v = -6\pi\eta Rv$ where $\eta = 1.8 \ 10^{-4}$ g/cm/s is the air viscosity. Solving $F_v = m \ dv/dt$ leads to the typical dynamics $v = v_o exp[-\alpha t]$, with $\alpha = 6\pi\eta R/m = 0.27 \ s^{-1}$. Three different regimes can be considered depending on the damping time $\alpha^{-1}$ compared to the 3 other parameters which are (i) the restitution coefficient $\epsilon = \Delta v/v$, (ii) the period $1/f$ of vibration and (iii) the time $T_r = L/(Af)$ needed by the particle to move from one wall to the other:

  (i)    if $f < \alpha$, the beads follow the periodic motion of the box due to the coupling via the air; this occurs when $f < \frac{1}{4}Hz$ in the present case. So this condition never occurs in the MiniTexus-5 experiments.

  (ii)   $\alpha L/(Af) < -\ln(\epsilon)$: viscous losses are negligible compared to grain-wall collision losses. Assuming $\epsilon = 0.98$, one gets that viscous damping is always larger than collision damping when $0.1 > L/(Af)$, *i.e.* except for experiment #9; so viscous damping is probably more important than collision damping in this experiment.

  (iii)  when $f > \alpha$ and $\alpha L/(Af) > -\ln\epsilon$, the viscous damping is larger than collision damping. However viscous damping can be considered as large only if $\alpha L/(Af)$ is larger than 1 because the particle speed will be quite different after the round trip to what it was before. When it is not, the losses can be calculated from a first order expansion and the particle speed $v$ decreases as $\delta v/v \approx -\alpha L/(Af)$ during a single wall-to-wall trip. This formula for loss is valid for most experiments in MiniTexus-5, except for experiments # 1 & 2 for which losses are larger, and experiment #9 for which collision losses are more important (see point ii).

At last, one shall mention that periodic forcing imposes a flow around the grain; this flow is mainly inertial except in a thin layer [14] whose thickness $\delta = (\nu/\omega)^{1/2}$ depends on the kinematic viscosity $\nu = \mu/\rho$ and on the frequency $f = \omega/(2\pi)$; as $\nu = 0.150cm^2/s$ for air, $0.2mm < \delta < 1.5mm$ in the MiniTexus-5 experiments. So, this effect may affect the dynamics of grain-grain collision when $f$ is small.

♣ Air effect on a cluster:

The air effect on a cluster of grains is a complicated problem due to ther existence of a viscous boundary layer whose thickness $\delta = (\nu/\omega)^{1/2}$ shall be compared to the mean grain-grain distance $l_g$. This leads to two different behaviours depending on whether $\delta < l_g$ or $\delta > l_g$.

But added to this, the drag force of the continuous flow itself depends on the cluster geometry and on the distance $l_g$; and both parameters lead to strong non linear behaviour. For instance, if the cluster is confined in a small volume the fluid can flow around it; this leads to a short-cut which makes the flow inside the porous quite negligible and the cluster geometry quite important. On the contrary, if the cluster expands over a whole section perpendicular to the main flow, the drag force can be computed using a porous flow model; it has been known from long that the porous permeability κ varies approximately as κ=R²Φ$^3$/45(1-Φ)² , where Φ is the cluster porosity [21]. This allows computing the pressure variation, its time dependency and the damping constant $\alpha_c$ of the cluster motion. As k varies quite non linearly with Φ the viscous drag is non linear too.

Any how some global approach can be performed when one can consider that the cluster does not interact directly with the pressure gauge but only via the surrounding fluid: in this case indeed, the fluid pressure near the pressure gauge shall increase and decrease periodically in phase with the acceleration. This explains the observed sinusoid variations in the case of the densest cell.

### *3.5. Effect of g-jitter:*

One can estimate the fluctuations of g to be less than $10^{-5}$g=$10^{-4}$m/s² in the range 100Hz-1Hz. So, the high frequency g-jitter, *i.e.* f>1 Hz is quite negligible compared to the mechanical noise of the vibrating device itself. The low-frequency component of g-jitter shall be even much smaller than $10^{-4}$m/s² since the rocket trajectory is compensated from it. However, this low frequency noise, which acts as a random force due to the viscous damping, shall force the cluster to perform a random walk. The time step of the walk is $\tau_w \approx$4s about, and the path length is likely smaller than 1mm.

Taking into account the typical flying time between the walls $\tau_f \approx$L/(Af), the random walk process becomes non negligible when $\tau_f$> $\tau_w$ only, which imposes Af/L<0.2s$^{-1}$ . According to Table 1, this is the case for experiments # 2-to-9.

### *3.6. Analysis of pressure data:*

#### ♣ Amplitude of the gauge signal:
We first remark that typical pressure signals have similar amplitudes in Figs. 2, 4 and 5, when looking at the same experiment # . This means that the signal is not too much perturb by the sensitivity of the gauge to the cell acceleration. Table 4 reports the typical expected pressure $\delta p_c$ due to a single collision of 1 bead (taken from Table 3), the typical signal amplitude S, (evaluated from Fig. 2), and their ratios S/ $\delta p_c$ for experiments #5-9.

| Experiment # | #5 | #6 | #7 | #8 | #9 |
|---|---|---|---|---|---|
| $\delta p_c$ per bead (from Table 3) | 0.19 Pa | 0.58 Pa | 1.9 Pa | 2.4 Pa | 4.8 Pa |
| S (Signal from Fig. 2) | 80mV=111 Pa | 0.2V=278Pa | 0.4V=556Pa | 1.5V=2kPa | 2V=2.8kPA |
| S/$\delta p_c$ | 580 | 480 | 290 | 833 | 580 |

*Table 4: Typical expected values of the pressure $\delta p_c$ due to a single collision (from Table 3), typical experimental signal amplitude S from the gauge, and their ratio S/$\delta p_c$ as a function of the different experiments. One finds that S/$\delta p_c$ is rather constant, but quite large.*

Table 4 demonstrates that the measured signal is 550±70 times larger than the predicted one; its relative standard deviation is 13%. One can think to few explanations for this rescaling factor: (i) it can be due to the large number of particles which could hit the gauge during the same sampling time; however, 550 is a much too large number in comparison to what it is expected, *cf.* Table 3. (ii) One can think also to a much shorter sampling time $\tau_s$, since $\tau_s$=1μs would explain the signal amplitude S; however, it would not explain the number of counts measured: the probability of measuring a 1μs signal with a 2kHz sampling rate is quite small, except if the number of collisions is multiplied by 500; so this explanation would require the signal to be much more erratic than it is observed experimentally.

So, one is faced to propose another explanation. Indeed we know that the transducer has a mechanical resonance at 40 kHz. So we are forced to propose that either this resonance or an other phenomenon such as electric charge,…, perturbs the detection and amplifies the mechanical response by a 550±70 factor about.

♣ Pressure statistics:
Pressure statistics are reported in Fig. 5. The probability distribution has been obtained by the direct counting of the number of data ranging between V and V+δV. However the used signals are those from Fig. 4 and not from Fig. 2, *i.e.* they are already corrected from the acceleration sensitivity. Also, according to what was discussed in the last paragraph, the scale of the pressure data has to be rescaled by a 550 factor to exhibit plausible typical pressures.

These distributions exhibit a large probability to get pressure around p=0, then a rather exponential tail which is characterised by a typical pressure $p_o$; $p_o$ depends on the experiment, it increases with the experiment #. The large probability to get the signal around p=0 is explained by the fact that the beads hits the gauge only during ¼ of a period. So it demonstrate the "supersonic" nature of the cell motion and of the "granular-gas" excitation.

*Interpretation of the experimental pressure distribution:* The exponential decrease is related to the excitation shape most likely: indeed if one follows this scheme, one expects the pressure signal to be proportional to the momentum variation; so, as the grain speed v is often much smaller than cell one $v_o(t)$ , the momentum change is mainly controlled by the cell speed $v_o(t)$ and the larger $v_o(t)$ the better the approximation. So, one expects (i) that the pressure signal is periodic, (ii) that it remains near p=0 till the gauge is not in contact with the "granular-gas", (iii) then it becomes maximum and (iv) decreases with time t as p≈ 1-sin ωt in the range t>0>π/2, till (v) p=0 when the gauge leaves contact with the granular-gas".

If one uses classical theory of gas, one can write the pressure signal as usual as:

$$P_z(t)=\sum_{\text{beads}} \rho(v_z,x_g,t) \, m(1+\varepsilon)[v_z(t)-v_o(t)]^2 \tag{5}$$

where $\rho(v_z,x_o,t)$ is the density of particles at location $x_g$ having the speed $v_z$ at time t. However, this equation holds true only in mean; in particular, it does not apply as is when the mean number of collisions becomes small, *i.e.* smaller than 1, which is the case for these experiments. So one has to perform a complete statistical analysis:

*Basis for a complete analysis of the fluctuation distribution:* Let consider that the beads are distributed uniformly and be $\rho(v)$ the probability distribution of speed v; consider also the 1d problem. As the distribution is uniform in space, any particle has the same probability $p_1=a/L$ to be in the volume a and $p_2=1-p_1=1-a/L$ to be in the complementary volume. Consider now a given duration $\tau$, and a particle having a speed v, this particle has the probability $v\tau/L$ to path through a given point and $1-v\tau/L$ not to path over this point. So one can write that the total distribution is described by:

$$W=1=\{(1-\int dv\ v\tau\rho(v)/L)+\int dv\ v\tau\rho(v)/L\ \}^{No}=\Sigma_n\ C^n_{N_o}\ (1-\int dv\ v\tau\rho(v)/L)^{No-n}\ (\int dv\ v\tau\rho(v)/L)^n$$

Where $C^n_{N_o}$ is the number of combinations of n objects taken among $N_o$. So considering the case $v\tau<<L$ and the limit $n<<N_o$, one gets:

$$(1-\int dv\ v\tau\rho(v)/L)^{N_o-n} \approx \exp\{-N\int dv\ v\tau\rho(v)/L\} \qquad \text{and} \qquad C^n_{N_o}\approx N^n/(n!)$$

$$W=1=\{(1-\Sigma_v v\tau\rho(v)/L)+\Sigma_v\ v\tau\rho(v)/L\}^{N_o}$$

So :

$$W=1=\exp\{-N_o\Sigma_v v\tau\rho(v)/L\}[\Sigma_n(\Sigma_v\ N_o v\tau\rho(v)/L)^n/(n!)] \qquad (6)$$

Indeed the term in the bracket [ ] is the expansion of $\exp\{N_o\Sigma_v v\tau\rho(v)/L\}$ . Furthermore, labelling $\underline{v}$ the mean speed of any particle and $b=L/N_o$ the mean "volume" per particle, the term $\exp\{-N_o\Sigma_v v\tau\rho(v)/L\}$ can be written as $\exp\{-\underline{v}\tau/b\}$ . So, each term with a given n in Eq. (6) can be viewed as the occurrence probability that n of the $N_o$ particles have passed through a given fixed z location during time $\tau$. It can be viewed as the probability $\Pi(n,v,\tau)$ that n particles hit a fixed boundary during the same lapse of time $\tau$. This term leads then to a pressure $p=n(1+\epsilon)mv$.

When the wall is mobile at speed $v_o(t)$, one has to replace v by $[v+v_o(t)]$ in Eq. (6) and each term with a given n leads to a pressure value $p=nm(1+\epsilon)[v+v_o(t)]$; its probability of occurrence $\Pi(n,v,v_o(t)),\tau_s)$ is:

$$\Pi(n,v,v_o(t),\tau_s)=\exp\{-N_o\Sigma_v[v+v_o(t)]\tau_s\rho(v)/L\}[(\Sigma_v\ N_o[v+v_o(t)]\tau_s\rho(v)/L)^n/(n!)] \qquad (7)$$

$$p=nm(1+\epsilon)[v+v_o(t)] \qquad (8)$$

In order to get the real pressure distribution, one has to sum the terms in Eq. (7) which correspond to the same pressure p according to Eq. (8). This will not be done here. Obviously it depends on the wall motion and on the probability distribution $\rho(v)$ of the grain speed.

However, one can perform an approximation when $v_o$ can be considered as much larger than v. In this case Eq. (8) leads to:

$$\Pi(n,v_o(t),\tau_s)=(N_o v_o(t)\tau_s/L)^n/(n!)\ \exp\{-N_o\ v_o(t)\ \tau_s/L\} \qquad (9)$$

which is the well-known Poisson distribution.

Eq. (9) leads to two different behaviours depending on the value of $\underline{n}$ = $N_o v_o(t)\tau_s/L$:

$\Rightarrow$ If $\underline{n}=N_o\,v_o(t)\,\tau_s/L <<1$, the $\exp\{\}$ term is approximately 1 and the terms $\underline{n}^n/n!$ decreases quickly with n; so each channel will get a signal S either equal to O or $S_1\sim$ p=m$(1+\varepsilon)v_o(t)$ with an occurrence probability $\Pi_o$ and $\Pi_1$ given by $\Pi_o$=1-$\underline{n}$ and $\Pi_1$=$\underline{n}$ . The mean signal <S> scales as <S> ~ $\underline{n}$m$(1+\varepsilon)v_o(t)$ = $N_o\tau_s$m$(1+\varepsilon)v_o(t)^2/$ L .

$\Rightarrow$ If $\underline{n}=N_o\,v_o(t)\,\tau_s/L >>1$, the $\exp\{\}$ term is small and the distribution is peaked around $\underline{n}$, with an approximate Gaussian shape. Neglecting the fluctuations lead to consider that each channel gives the same signal S~p= $\underline{n}$m$(1+\varepsilon)v_o(t)$ = $N_o\tau_s$m$(1+\varepsilon)v_o(t)^2/$ L .

So, both conditions, *i.e.* $\underline{n}$<1 or $\underline{n}$>1, do not lead to the same distribution law, even if it results in the same mean.

*End of interpretation of the experimental pressure distribution:* One expects from Table 3, that experiments # 1-5 are concerned with less than 1 collision per sampling time whereas experiments # 6-9 are concerned with more than 1 collision per sampling time. So one expects that recorded pressure scales as p~m$(1+\varepsilon)(v+v_o)$ in the first case and as p~m$(1+\varepsilon)(v+v_o)^2$ in mean in the second case (indeed, the second [v-$v_o$] factor takes into account the mean particle number per sampling). So one can understand the p~$(A\omega)^{3/2}$ experimental fit as resulting from a compromise between the two previous scalings $A\omega$ & $(A\omega)^2$.

One shall remark also that neither the real mean speed v of the particle nor its dispersion shall have some important effect since the pressure is dominated by the speed of impact which is mainly the wall speed in the case of "supersonic excitation", which is just the case when the pressure is large.

However the grain speed distribution may have some importance at small pressure, when the wall speed is slow at the beginning of the "exponential" tail.

The last part of the exponential tail corresponds to large pressure; then this part of the dispersion cannot come from the speed distribution, since this one is mainly limited by the maximum speed of the cell. Then this exponential tail reflects the fluctuations of the amplification factor, *i.e.* 550±70, which is needed to introduce to understand the amplitude of the signal.

After all, it is worth recalling (i) that the large number of counts which occur with p=0 in Fig. 5 is due to the large lapse of time of each period during which no collision occurs due to the "supersonic motion" of the cell, and (ii) that the beginning of the "exponential tail" reflects the periodic decrease of cell speed. In such conditions, the Gaussian distribution which is expected for classical gas cannot be observed in a granular gas excited periodically along the same direction as it it is excited.

## 3.6.  Other questions and conclusion:

♣ Effect of grain-grain collision

One of the main question is to understand why the typical grain speed is much smaller than the cell one: indeed 1d single-particle model predicts $\underline{v}>A\omega$, for all systems, *i.e.* even if the restitution coefficient r is small. But video analysis demonstrates the contrary, at least when A is large enough. On the other hand viscous damping is expected to be quite small under these conditions, *cf.* Table 3; furthermore, effect of viscous damping is expected to decrease when the typical speed increases. So one is faced to a dilemma: why is the measured typical speed much smaller than the expected one. Indeed, air turbulence can be invoked, but the Reynolds number $R_e$ of the flow remains small, *i.e.* $R_e = 2v_oR/\nu = 4\pi AfR/\nu = 1.2Af<15$, since $\nu=0.15$cm²/s, $2R\approx0.3$; so this contradicts this explanation.

One of the main difference between 1d and 3d "granular-gas" behaviours comes from the efficient redistribution of speed direction due to grain-grain collisions in 3d; this amplifies the process of randomisation: Even when the restitution coefficient r, $r^2=\varepsilon$, of the collision is equal to 1, the direction of the speeds after a collision depends on an extra parameter, called the impact parameter, in 2d and in 3d. This parameter is defined as the ratio between the minimum distance of the two trajectories divided by the grain diameter. So even for gas with small density, the number of collisions can be large enough to destroy motion with resonant condition if it exists. This becomes more efficient when the mean free path $l_c$ becomes of order L , as it is in the above experimental case. Furthermore, one has to introduce rotation kinetic energy most likely, since solid friction shall transform some part of the translation kinetic energy into the rotation one. In order to predict the real mean speed, one should know the typical function describing the speed distribution. This is not known unfortunately at present.

However, in order to predict it, one can perhaps assume that the speed distribution $F(v_x,v_y,v_z)$ of the particles is completely randomised by the 3d grain-grain collision process; so, if one admits also that the speed distribution in one direction is independent of the other ones, one can develop an analysis similar to the one proposed by Maxwell [21] and finds that $p(v_x,v_y, v_z)=\exp\{-m(v_x^2+v_y^2+ v_z^2)/(kT)\}$.

However, we think this hypothesis is likely not true in the present case since the speed distribution in the x and y directions shall depend likely on $v_z$ . So, at this stage of the present research I have not found any precise argument in favour of a mean speed $\underline{v}$ of particle smaller than $A\omega$. But I do think it is likely related to the 3d dynamics, instead of the 1d one.

♣ Conclusion

Data analysis of Minitexus 5 experiments has been performed in the case of the loose sample, starting from the video observation that the cell moves faster than the grain. This has allowed to postulate a supersonic excitation.

In a previous paper [6], it was found that the granular-pressure distribution scales as $p\sim(A\omega)^{3/2}$. One of the main result of this paper is that this scaling results most likely from a crossover between two dynamics regime: the first regime is obtained at small Aω corresponds to at most one collision during each sampling time (exp#1-5) ; the second regime occurs at large Aω and corresponds to a multiple-collision (exp#6-9). In previous papers [6,7,10], we have attributed this behaviour to the dependence of the restitution coefficient ε on the speed [6], or to a finite size effect [10]; this is also possible, and more work is required to conclude. Another important result is the qualitative interpretation of the statistics of the pressure fluctuations: the large number of counts which occur near p≈0 is due to the supersonic regime of excitation, which disconnects the gauge from the granular gas during a long period of time. The fastest collisions, leading to the largest pressures, occur when the gauge hits periodically the "granular-gas" phase. So, the beginning of the distribution tail is related to the cell speed distribution; but it is thought that the end of the exponential tail is related to the statistics of the noise produced by the gauge. So, it results from the "supersonic" gauge motion that the pressure distribution is not Gaussian-shaped when pressure is measured in the direction of vibration; however, it may be Gaussian-shaped when pressure is measured with a motionless gauge and perpendicularly to the excitation. So even if pressure p is a much easier quantity to measure experimentally than granular temperature it does not always behave the same, even for a spatially homogeneous density.

Effect of air viscosity and of g-jitter have been quantified. Analogy with classical gas and with single particle problem have been drawn and discussed. These have lead to a second important result: despite what one can think at first sight, a shaken granular material does not behave as classical gases in classical regime, even in weightlessness condition. Indeed, if it is too dense it forms a cluster spontaneously, which reduces the density of the "granular-gas" phase surrounding the cluster; this "granular-gas" is in a Knudsen regime, with a mean free path larger than the cell size; its excitation is of the "supersonic" kind. This effect results most likely from the non-elasticity of grain-grain collisions and from the 3d nature of the system.

Since the pressure data of these experiments shall be rescaled by a factor 550±70 to get estimated values, we think that more works are required to confirm these data, using first a more confident pressure gauge; these experiments are currently being performed in the CNES Airbus A300. We do not understand the origin of this factor, but we suppose it is linked to the 40kHz mechanical resonance of the gauge itself.

Also much more works are needed to investigate both the gas-cluster transition and the cluster mechanics itself in weightlessness. This study cannot be performed in the Airbus due to its large g-jitter, *i.e.* $10^{-2}$g=0.1m/s², which leads to an estimated free-flying time $\tau_c$=0.3s for the cluster in the cavity. Increasing the air- pressure and density is also a major goal in order to study the mechanics of multiphase systems in weightlessness, or to replace the air by some viscous liquid.

***In conclusion,*** we have reported a 3D experiment of a granular medium fluidised by sinusoidal vibrations in low gravity environment. When the density of the granular medium is increased, we clearly show that an ensemble of particles in erratic motion interacting only through inelastic collisions spontaneously generates the formation of a motionless dense cluster.

*Acknowledgements:* I must thank D. Beysens & Y. Garrabos without whom this experiment could not have been performed, who participated to the definition of the experimental set-up and defended this project. Collaboration with D. Beysens, Y. Garrabos, C. Chabot-Lecoutre and R. Wunenburger has been quite appreciated during the whole launching campaign. Collaboration with E. Falcon and S. Fauve has been also quite appreciated; in particular for the statistical treatment of the pressure signals and for clever ideas.

CNES is thanked for funding and ESA for providing the experiment and the flight opportunity; *Mini-Texus* 5 sounding rocket is a program of E.S.A.; MiniTexus 5 was launched from Esrange on February 1998. The bronze spheres have been provided by Makin Metal Powders Ltd. The experiment module has been constructed by D.A.S.A. (Germany**),** Ferrari (Italy), and Techno System (Italy). We gratefully acknowledge the Texus team for its kind technical assistance.

## References

[1] J.C. Worms, H. de Maximy, réalisation Ch. Bargues, "Une loi réputée simple: pV=nRT"; (Planet 6 & CNES ed., Planet 6 c/o J.C. Worms, 12 rue de l'Espérance, 67400 Illkirch, France , 1993)

[2] T. Pöschel & S. Luding eds., *Granular gases*, (Lecture Notes in physics series, Springer, Berlin, 2001), and refs. there in.

[3] Y. Du, H. Li & L.P. Kadanoff, "Breakdown of Hydrodynamics in a one dimensional system of inelastic particles", Phys. Rev. Lett. **74**, 1268, (1995)

[4] I. Goldhirsch, "Granular gases: Probing the boundaries of hydrodynamics", in *Granular gases*, (Lecture Notes in physics series, Springer, Berlin, 2001), pp. 79-99, and refs. there in.

[5] A. Goldshtein, A. Alexeev, & M. Shapiro, "Resonance Oscillations in granular gases", in *Granular gases*, (Lecture Notes in physics series, Springer, Berlin, 2001), pp. 266-277

[6] E. Falcon, R. Wunenburger, P. Évesque, S. Fauve, C. Chabot, Y. Garrabos and D. Beysens, *Phys. Rev. Lett.* **83**, 440-443 (1999)

[7] P. Évesque, E. Falcon, R. Wunenburger, S. Fauve, C. Lecoutre-Chabot, Y. Garrabos and D. Beysens, "Gas-cluster transition of Granular Matter under Vibration in Microgravity" , in Proc. of the "First international Symposium on Microgravity Research & Applications in Physical Science and Biotechnology", 10-15 Sept 2000, Sorrento, Italy, pp., 829-834 (B. Schürmann ESA ed. , 2001)

[8] P. Goldreich and S. Tremaine, *Ann. Rev. Astron. Astrophys.* **20**, 249 (1982).

[9] A. Kudrolli, M. Wolpert and J. P. Gollub, *Phys. Rev. Lett.* **78**, 1383 (1997).

[10] P. Evesque, "The thermodynamics of a single bead in a vibrated container", *poudres & grains* **12**, 17-42, (2001), http://prunier.mss.ecp.fr/poudres&grains/poudres-index.htm

[11] J.O. Hirschfelder, C.F. Curtis & R.B. Bird, *Molecular theory of gases and liquids*, (John Wiley & sons eds, New York, 1967), pp. 694-704 and see problems related to Knudsen gases

[12] P. Evesque, "1-d granular gas with little dissipation in 0-g : A comment on "Resonance oscillations in Granular gases"", *poudres & grains* **12 (3)**, 50-59, (2001), http://prunier.mss.ecp.fr /poudres&grains/ poudres-index.htm

[13] E. Falcon, C. Laroche, S. Fauve and C. Coste, *Eur. Phys.J.* B **3**, 45 (1998).

[14] L. Landau & E. Lifschitz, *Mécanique des Fluides (cours de Physique tome 6)*, (Mir, 1989), pp. 123, 356-357, 429-433, 437-440, 448-504

[15] M. Greenspan:"Transmission of soundwaves in gases at very low pressure", in *Physical Acoustics*, tome 2-A, pp. 1-45, (W.P. Mason ed., Academic press, 1965).

[16] R.K. Cook, M. Greenspan & M.C. Thompson Jr., J. Acoust. Soc. Am. **25**,192, (1953)

[17] A.A. Ivanova, V.G. Kozlov, D.V. Liubimov, T.P. Liubimova, "Structure of averaged flow driven by a vibrating body with a large-curvature edge", *Fluid Dynamics* **33**, 659-666 (1998)

[18] Lord Rayleigh, "On the circulation of air in the Kundt's tube, and on some allied acoustical problems", *Phil. Trans. Roy. Soc. London Ser.* **A 175**, 1, (1883)

[19] H. Schlichting, *Boundary Layer Theory*, Mac Graw Hill, (1968)

[20] J. Lighthill, "Acoustic streaming", *J. Sound Vibrat.* **61** , 39l, (1978)

[21] G. Bruhat, *Cours de Physique Générale: Thermodynamique*, pp. 460-467, (Masson, Paris, 1968)

# Feed-back from Readers :

## *Discussion, Comments and Answers*

---

## From *poudres & grains* articles:

On *Poudres & grains* **12**, 17-42 (2001): In Figs 8,9,10, l3, Log means Neperian Logarithm. L/2 is the half of box length in Figs. 11, 12 & 13.

*From a remark by Y. Garrabos:* The rapid decrease of V/(bω) *vs.* ln(2b/L), with the increase of b/L, is due to the synchronisation of the bead motion on the excitation, leading to regular impact at a frequency equal to half the excitation frequency ν {this occurs around V/(bω)=4 and ln(2b/L)=-2.53, *i.e.* and corresponds to a one way (≈L) which takes at time 1/ν } or at ν {this occurs around V/(bω)=4 and ln(2b/L)=-1.84, *i.e.* round-trip (≈2L) takes 1/ν }. This last synchronisation has been observed in recent Airbus result.

P.E.

On *Poudres & grains* **12**, 60-82 (2001): *Remark by L. Ponson, P. Burban, H. Bellenger & P. Jean* : According to the data, $l_c$=L/5 to L/6 for the less dense sample.

*Answer:* Indeed, the mean free path $l_c$ is smaller than the cell size L; however, it remains of the same order; so, this does not invalidate the qualitative findings and the theoretical discussion about the Knudsen regime; moreover, one expects that a decrease of $l_c$/L generates an increase of loss, this value of $l_c$/L may help understanding why the mean bead speed is so slow in the MniTexus experiment, compared to what one shall expect from 1-bead simulations.

P.E.

On *Poudres & grains* **12**, 107-114 (2001): lines 6-7 of $2^{nd}$ paragraph of section 3: the liquid flow exists but is small in the limit of a small toroidal section of radius $R_s$, *i.e.* proportional to $R_s/R_t$, due to preservation rules as stated in the introduction. In this limit this affects little the results. Furthermore, it is known that ripples formation occurs above the threshold at which the Stokes boundary layer is destabilised, *i.e.* when the Reynolds number $R_{eδ}$= bΩδ/ν=2b/δ is larger than 100-to-500, for which δ= √(2ν/Ω) is the viscous boundary layer thickness and b is the relative amplitude of the flow motion; so, in the present case it scales as $R_{eδ}$ =2$α_m R_s$/δ , for which $α_m$ is the amplitude of rotation. It has been found also that ripple wavelength λ is about 12δ at threshold. Limit of Stability of the Stokes boundary layer can be found in V.G. Kozlov, Stability of periodic motion of fluid in a planar channel // Fluid Dynamics, 1979, vol. 14, no. 6, 904–908 and in Stability of high-frequency oscillating flow in channels // Heat Transfer – Soviet Research, 1991, v. 23, no. 7, 968–976.

P.E.

## About published articles from other Reviews

About *Phys.Rev. Lett.* **84**, (2000) 5126, and *C.R. Physique* **3** (2002) 217-227, by J. Duran: Formation of such Ripples and its domain of existence has been studied under sinusoid vertical forcing by V.G. Kozlov, A. Ivanova and P. Evesque (Europhys. Lett. **42**,413-418 (1998)) with a viscous fluid instead of air. Similar heap formation has been found; it was also found .

P.E.

The electronic arXiv.org version of this paper has been settled during a stay at the Kavli Institute of Theoretical Physics of the University of California at Santa Barbara (KITP-UCSB), in june 2005, supported in part by the National Science Fundation under Grant n° PHY99-07949.

*Poudres & Grains* can be found at :
http://www.mssmat.ecp.fr/rubrique.php3?id_rubrique=402